\documentclass[Journal,onecolumn]{IEEEtran}
\usepackage{amsmath}
\usepackage{theorem}
\usepackage{amsmath}
\usepackage{mathrsfs}
\usepackage{accents}
\usepackage{setspace}
\usepackage{url}
\usepackage{algorithm}
\usepackage{algorithmic}
\usepackage{epsfig}
\usepackage{amsfonts}
\usepackage{amssymb}
\usepackage{mathrsfs}
\usepackage{euscript}
\usepackage{graphicx}
\usepackage{multirow}
\usepackage{cite}
\usepackage{amsmath}
\usepackage{theorem}
\usepackage{amsmath}
\usepackage{mathrsfs}
\usepackage{accents}
\usepackage{url}
\usepackage{tabularx}
\usepackage{float}
\usepackage{amsmath}
\usepackage{algorithm}
\usepackage{algorithm}
\usepackage{algorithmic}
\usepackage{epsfig}
\usepackage{amsfonts}
\usepackage{amssymb}
\usepackage{mathrsfs}
\usepackage{euscript}
\usepackage{graphicx}
\usepackage{multirow}
\usepackage{cite}
\usepackage{placeins}
\usepackage{caption}
\usepackage{color,soul}
\usepackage{soul}
\usepackage{subcaption}
\usepackage{lipsum} 
\usepackage{filecontents} 
\newtheorem{theorem}{Theorem}
\makeatletter

\def\subsubsection{\@startsection{subsubsection}
                                 {3}
                                 {\z@}
                                 {0ex plus 0.1ex minus 0.1ex}
                                 {0ex}
                                 {\normalfont\normalsize\itshape}}
\makeatother

\usepackage[T1]{fontenc}
\usepackage[nodayofweek,level]{datetime}
\newcommand{\mydate}{\formatdate{6}{6}{2017}}

\graphicspath{{./}{figures/}}

\ifCLASSINFOpdf
\else
\fi
\begin{document}

\begin{titlepage}

\begin{tabular}{l        r}

\includegraphics[bb=20bp 00bp 500bp 450bp,clip,scale=0.3]{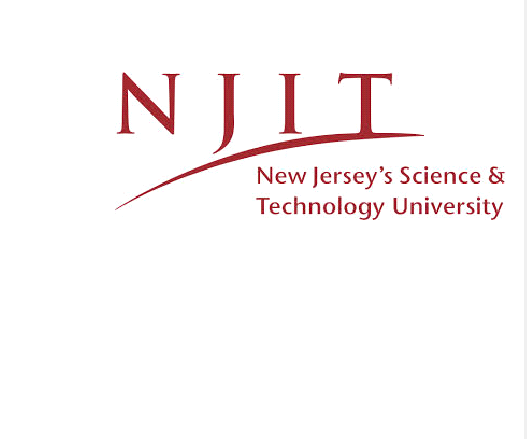} \hspace{6cm} & \includegraphics[bb=0bp -200bp 500bp 550bp,clip,scale=0.2]{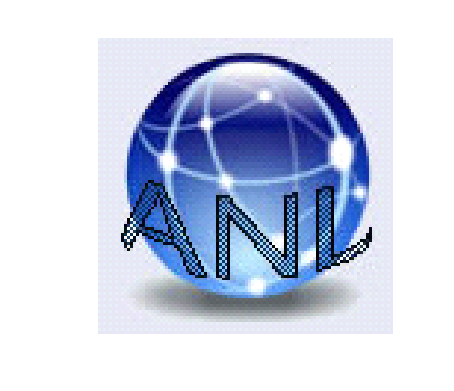}

\end{tabular}

\begin{center}

\textsc{\LARGE Hierarchical Capacity Provisioning for Fog Computing}\\[1.5cm]

{\Large \textsc{Abbas Kiani}}\\ 
{\Large \textsc{Nirwan Ansari}}\\ 
{\Large \textsc{Abdallah Khreishah}}\\
[2cm]

{}
{\textsc{TR-ANL-2017-005}\\
\large \mydate} \\[3cm]

{\textsc{Advanced Networking Laboratory}}\\
{\textsc{Department of Electrical and Computer Engineering}}\\
{\textsc{New Jersy Institute of Technology}}\\[1.5cm]
\vfill

\end{center}

\end{titlepage}

\begin{spacing}{2}
\begin{abstract}
The concept of fog computing is centered around providing computation resources at the edge of network, thereby reducing the latency and improving the quality of service. However, it is still desirable to investigate how and where at the edge of the network the computation capacity should be provisioned. To this end, we propose a hierarchical capacity provisioning scheme. In particular, we consider a two-tier network architecture consisting of shallow and deep cloudlets and explore the benefits of hierarchical capacity based on queueing analysis. Moreover, we explore two different network scenarios in which the network delay between the two tiers is negligible as well as the case that the deep cloudlet is located somewhere
deeper in the network and thus the delay is significant. More importantly, we model the first network delay scenario with bufferless shallow cloudlets as well as the second scenario with finite-size buffer shallow cloudlets, and formulate an optimization problem for each model. We also use stochastic ordering to solve the optimization problem formulated for the first model and an upper bound based technique is proposed for the second model. The performance of the proposed scheme is evaluated via simulations in which we show the accuracy of the proposed upper bound technique as well as the queue length estimation approach for both randomly generated input and real trace data.
\end{abstract}

\section{Introduction}\label{sec:Introduction}
\IEEEPARstart{T}{he} paradigm of edge computing has been recently introduced to push the computing resources away from the centralized nodes to the edge of the network.
The edge computing concept aims at optimizing the cloud computing networks by reducing the communication bandwidth requirement between the sources of data and the data centers. In fact, edge computing pushes the processing power and intelligence directly to the devices and each device in the network can play its own role in processing the data.
While the edge computing architectures are developed in different ways and are designed with different names, the main idea is to reduce the latency and improve the Quality of Service (QoS). For example, the idea of cloudlet as a trusted, resource-rich computer which is available for use by nearby mobile devices was first introduced in~\cite{satyanarayanan2009case}, and further developed by a research team at Carnegie Mellon University~\cite{clinch2012close,satyanarayanan2013role,ha2015adaptive}.
A cloudlet can be considered as a small-scale cloud data center (known as a data center in a box or mobile micro-cloud) which supports resource-intensive mobile applications by providing computing resources at the edge of the network~\cite{satyanarayanan2009case}.
Three years after the introduction of the cloudlet concept, the paradigm of fog computing~\cite{bonomi2011connected} was introduced by Cisco as a multi-tiered architecture consisting of the device, fog platform and a data center to support the requirements of Internet of Things (IoTs)~\cite{bonomi2012fog}.
Fog computing (also known as fogging) is basically a decentralized computing infrastructure that extends cloud computing network by distributing the computing as well as storage resources in efficient places (fog nodes) between where the data is created and the cloud~\cite{bonomi2011connected}.

In parallel with the development of both fog computing and the cloudlet concepts, the idea of Mobile Edge Computing (MEC) has been standardized by an Industry Specification Group (ISG) lunched by the European Telecommunications Standards Institute (ETSI)~\cite{hu2015mobile}. MEC is recognized as one of the key emerging technologies for 5G networks and aims at providing computing capabilities within the Radio Access Network (RAN) and in proximity of mobile users~\cite{hu2015mobile,sun2017green}. Smart mobility, smart cities, and location-based services are named as the potential IoT applications of MEC~\cite{corcoran2016mobile,sun2}.

In the past few years, a large and cohesive body of work investigated the major limitations of Mobile Cloud Computing (MCC), e.g., the radio access associated energy consumption of mobile devices~\cite{7846741} and the latency experienced over Wide Area Network (WAN). Based on that, the researchers came up with a variety of policies and algorithms. A cloudlet network planning approach for mobile access networks is introduced in~\cite{ceselli2015cloudlet} which optimally places the cloudlet facilities among a given set of available sites and then assigns a set of access points to the cloudlets by taking into consideration of the user mobility.
Recently, the opportunities and challenges of edge computing in the networking context of IoT is summarized in~\cite{chiang2016fog}. The authors in~\cite{jutila2016adaptive} investigate adaptive edge computing solutions for IoT networking which aim at optimizing traffic flows and network resources. The state of the art of edge computing and its applications in IoT is also explored in~\cite{gonzalez2016fog}. A hybrid architecture that harnesses the synergies between  edge caching and C-RAN is proposed in~\cite{tandon2016harnessing}. Moreover, a three levels cloudlet architecture is designed in~\cite{kiani2016towards} in which the authors proposed a two time scale approach for allocating the computing and communication resources to satisfy the users' QoS.

To shed some light on the idea of this paper, let's consider distributed CCTV video cameras as a potential application of edge computing.
For example, more than 400 CCTV video cameras are distributed over the state of New Jersey and they are generating a huge amount of video data each day. These data have to be processed and stored for different applications such as traffic congestion mitigation strategies.
However, sending all of these data to a backend system such as Traffic Management Centers (TMC), which is equipped with computational and storage capabilities, is not practical due to two main reasons: 1) The opportunity to process video data and act on the processed data might be gone after the time it takes to send data all the way to TMC over the backhaul network. 2) Continuously capturing video on the cameras poses a permanent stress on the network paths to the centralized controller.
One simple solution to mitigate the congestion on the backhaul network may offer buffering data at the intermediate network nodes for later transmission. However, this solution is not useful because cameras are capturing videos 24/7 and there will never be a future time when the backhaul network is not overwhelmed. Another solution towards this problem can be a distributed edge computing network architecture by leveraging the concept of the cloudlets. In such a distributed network architecture, each camera itself as well as the aggregation nodes in the network such as the network hubs and routers are all the potential sites to install the cloudlets.
Therefore, two important questions must be answered about such a distributed edge computing architecture: 1) Should we consider a flat or hierarchical design?
2) What is the size of each cloudlet, i.e., how much capacity should be provisioned at each cloudlet location?
To this end, the current study aims to address the aforementioned issue by proposing a hierarchical capacity provisioning scheme.
It is worth mentioning that this study is focused on the capacity provisioning as a network planing problem, and we do not take into consideration of the workload allocation problem which is a well studied issue in state of the art papers\cite{sun2017latency,fan2017energy,sun2017adaptive}. In fact, the idea here is to efficiently provision a total capacity budget at the edge while the distribution of the computation workload at different locations is given.

\textbf{Contributions:} We have made two major contributions. 1) We propose a hierarchical capacity provisioning scheme by considering a 2-tier edge computing network architecture consisting of shallow and deep cloudlets.
2) We investigate two different network scenarios based on accurate queueing analysis. In particular, we study the
case that the network delay between the shallow cloudlets
and the deep cloudlet is negligible as well as the case in which the deep cloudlet is located
somewhere deeper in the network, and thus the network delay
between the shallow cloudlets and the deep cloudlet matters. We also formulate optimization problems for each case and investigate the solution to each problem by using stochastic ordering and optimization algorithms.

The rest of the paper is organized as follows. Section~\ref{Sec:Model} describes the system model and problem formulation. We propose our hierarchical capacity provisioning scheme and the corresponding optimization problems in Section~\ref{Sec:profit}. Finally, Sections~\ref{sec:simulations} and~\ref{sec:conclude} present numerical results and conclude the paper, respectively.

\section{System Model and Problem Formulation}\label{Sec:Model}
We consider a fog computing network consisting of $M$ shallow cloudlets as the first tier of a two-tier hierarchical fog computing architecture.  Accordingly, the second tier of fog computing nodes called the deep cloudlet is connected to all the shallow cloudlets. Therefore, we assume that each shallow cloudlet can cooperatively manage its incoming workload with the deep cloudlet. That is, the peak computing load at a shallow cloudlet can be forwarded to the deep cloudlet. As a practical case, we consider a distributed edge video processing environment shown in Fig.~\ref{fig:1}. However, the proposed hierarchical capacity provisioning framework in this paper is not limited to only this example and it is applicable to all similar edge computing architectures. As depicted in this example, the shallow cloudlets are co-located with CCTV cameras and the deep cloudlet is installed at an aggregation switch.  Moreover, in order to leverage the resource-rich facilities, the deep cloudlet is connected to the cloud via fibers. However, our focus in this paper is on the capacity provisioning at the edge, i.e., the shallow and deep cloudlets.

\begin{figure}\label{fig:1}
\epsfig{file=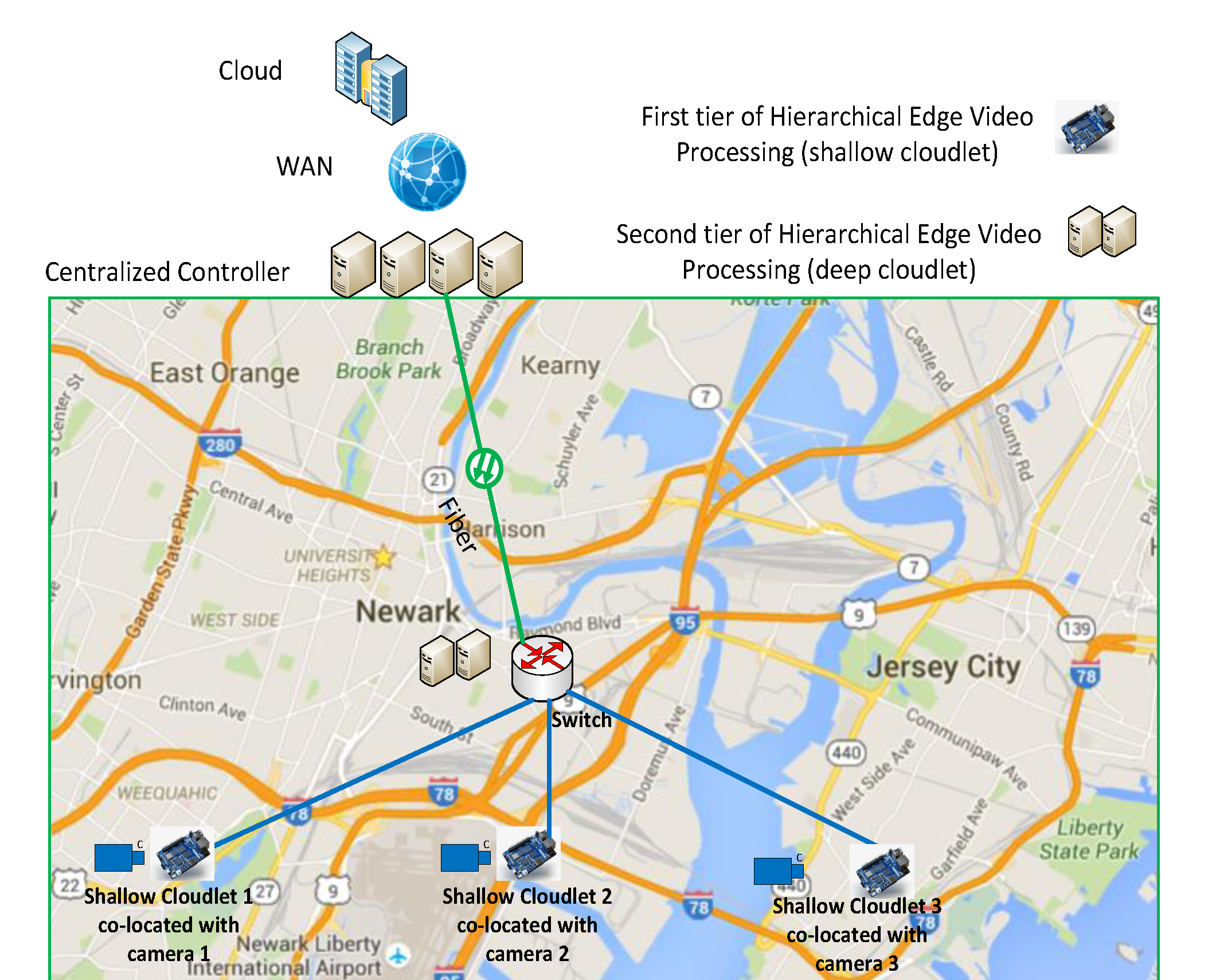,width=1.1\linewidth,clip=}
\caption{System model.}
\label{fig:1}
\end{figure}

\begin{figure*}\label{fig:7}
\epsfig{file=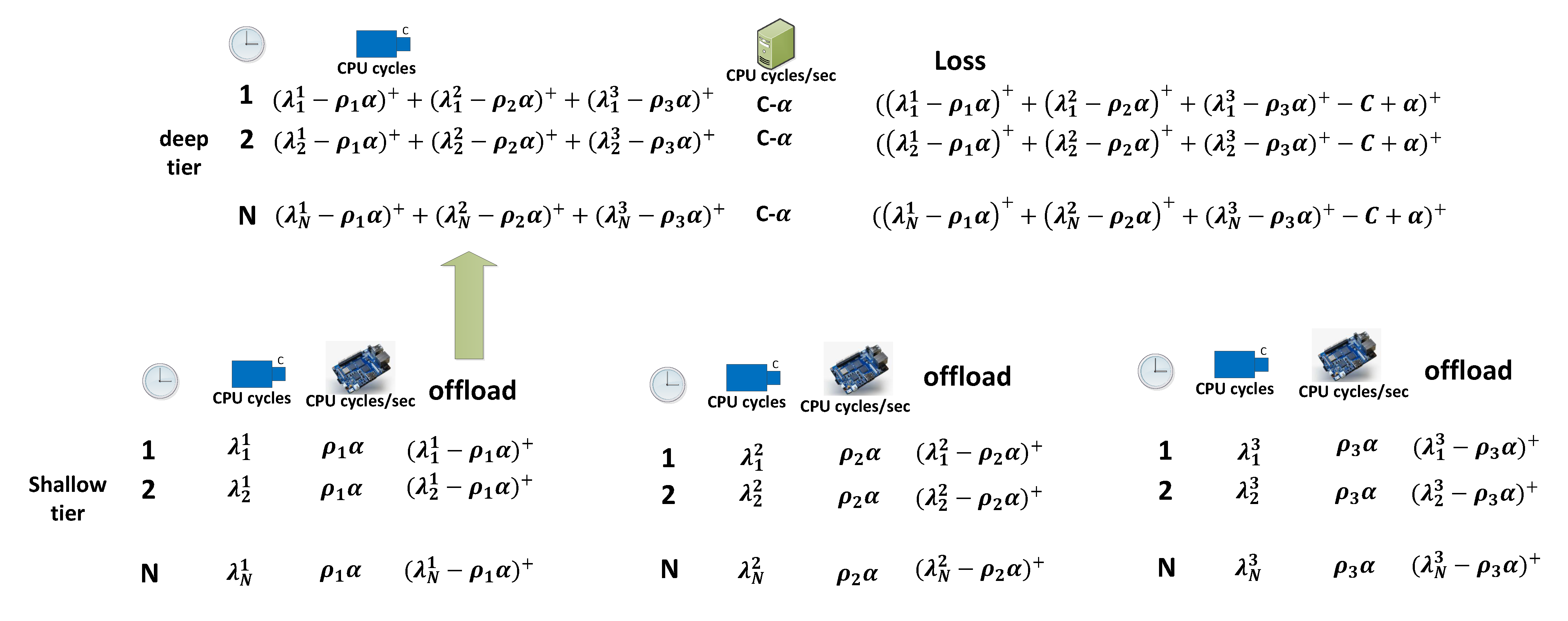,width=1\linewidth,clip=}
\caption{System model for bufferless shallow cloudlets.}
\label{fig:7}
\end{figure*}

We assume that the amount of edge computing workload at each shallow cloudlet at a given time follows a general distribution. We also assume that $C$ is the total capacity budget to be provisioned at the edge where a portion $\alpha$ of the capacity is provisioned at the shallow cloudlets and $C-\alpha$ at the deep cloudlet. Both the workload and the capacity are measured in CPU cycles.
We use CPU cycles to measure the workload since it has been widely used in the literature to measure the computation requirements of the computing tasks~\cite{cuervo2010maui}. Accordingly, to be consistent with the workload unit, we use CPU cycles per second as the unit of the computing capacity.
Moreover, we consider a finite size queuing system at each cloudlet location where all the queuing systems are modeled as a discrete-time fluid system.
In particular, at each time $n$, the queuing system at shallow cloudlet $i$ consists of a server with constant rate $\rho_i\alpha$ and a fluid input $\lambda_n^i$  which is assumed to be ergodic and stationary. We assume that $\lambda_n^i$'s are independent but have a common distribution and $E(\lambda_n^i)=\overline{\lambda}_i$. The normalized coefficient $\rho_i$ is also defined as $\rho_i=\frac{\overline{\lambda}_i}{\sum_{i=1}^M\overline{\lambda}_i}$. The system is assumed to be stable, i.e., $\sum_{i=1}^M\overline{\lambda}_i\leq C$.

\section{Capacity provisioning}\label{Sec:profit}
We investigate two different network scenarios for the proposed system model. In particular, we first investigate the case that the network delay between the shallow cloudlets and the deep cloudlet is negligible. In the second scenario, we consider the case in which the deep cloudlet is located somewhere deeper in the network, and thus the network delay between the shallow cloudlets and the deep cloudlet is significant.

\subsection{Bufferless shallow cloudlets}
We first investigate a network model in which the network delay between shallow cloudlets and the deep cloudlet is negligible.
As shown in Fig.~\ref{fig:7}, for such a network, we consider a buffer of size zero at each shallow cloudlet.
Note that going from a flat architecture consisting of only shallow cloudlets to a hierarchical architecture with both the shallow cloudlets and the deep cloudlet, we take a portion of the capacity of the shallow cloudlets and allocate it to the deep cloudlet. Such a hierarchical capacity provisioning model is fair only if one unit of the capacity at a shallow cloudlet results in the same delay as compared to that at the deep cloudlet. Therefore, when the network delay is negligible, this fairness requirement is satisfied with bufferless shallow cloudlets since the deep cloudlet is assumed to be bufferless too. In other words, considering buffers at the shallow cloudlets while the deep cloudlet is bufferless is not a fair assumption from the perspective of the proposed capacity provisioning model.
At each time $n$, the amount of the computing workload forwarded to the deep cloudlet is equal to $\sum_{i=1}^{M}(\lambda_n^i-\rho_i\alpha)^+$ where $(x)^+=max(x,0)$.
Accordingly, the queuing system of the deep cloudlet can be modeled  as a discrete-time fluid system consisting of a single server of constant rate $C-\alpha$ and a fluid input $\sum_{i=1}^{M}(\lambda_n^i-\rho_i\alpha)^+$.
At time $n$, the total amount of fluid loss in the system can be established as $(\sum_{i=1}^{M}(\lambda_n^i-\rho_i\alpha)^+-(C-\alpha))^+$.
The average fluid loss in the system is calculated as
\begin{eqnarray}\label{equ1}
\overline{L}_{bl}(\alpha)=\lim_{N\to\infty}\frac{\sum_{n=1}^{N}(\sum_{i=1}^{M}(\lambda_n^i-\rho_i\alpha)^+-(C-\alpha))^+}{N}=
E((\sum_{i=1}^{M}(\lambda_n^i-\rho_i\alpha)^+-(C-\alpha))^+)
\end{eqnarray}
where the second equality is due to the ergodicity assumption. Note that the focus of this paper is on proposing a network capacity planning framework rather than a workload placement algorithm. Therefore, to achieve an optimum capacity provisioning, we propose to solve the following optimization problem
\begin{eqnarray}
\underset{\alpha}{\text{minimize}}~\overline{L}_{bl}(\alpha)\label{equ2}
\end{eqnarray}
\begin{eqnarray}
s.t.~C_1:\sum_{i=1}^{M}E(\lambda_n^i-\rho_i\alpha)^+\leq C-\alpha\nonumber
\end{eqnarray}
\begin{eqnarray}
C_2: 0\leq\alpha\leq C\nonumber
\end{eqnarray}
where the objective is to minimize the average fluid loss and constraint $C_1$ is necessary for stabilizing the queue at the deep cloudlet. The following theorem provides an optimal solution to problem~(\ref{equ2}).

\begin{theorem}\label{theorem1}
The optimal solution to optimization problem (\ref{equ2}) is achieved when $\alpha=0$, i.e., when all the computing capacity is provisioned at the deep cloudlet.
\end{theorem} \begin{IEEEproof}
To prove Theorem~\ref{theorem1}, we need to show that $\overline{L}_{bl}(\alpha)$ is an strictly increasing function with respect to $\alpha$.
After some simple algebraic manipulation on $\overline{L}_{bl}(\alpha)$, we have,
\begin{equation}\label{equ3}
\begin{split}
\overline{L}_{bl}(\alpha)=E((\sum_{i=1}^{M}\max(\lambda_n^i,\rho_i\alpha)-C)^{+})
\end{split}
\end{equation}
Function $\overline{L}_{bl}(\alpha)$ is proven to be strictly increasing if we can show that $\overline{L}_{bl}(\alpha_h)<\overline{L}_{bl}(\alpha_k)$ for all $\alpha_h<\alpha_k$, where $0\leq\alpha_h,\alpha_k\leq C$. Consider two random variables $X_n=\sum_{i=1}^{M}\max(\lambda_n^i,\rho_i\alpha_h)$ and $Y_n=\sum_{i=1}^{M}\max(\lambda_n^i,\rho_i\alpha_k)$. If $X_n$ and $Y_n$ satisfy the stop-loss order, written as $X_n<_{sl}Y_n$, then $\overline{L}_{bl}(\alpha_h)<\overline{L}_{bl}(\alpha_k)$ for all $C$. In addition, the stop-loss order is maintained under the summation of independent random variables. Therefore, if random variable $\max(\lambda_n^i,\rho_i\alpha_h)$ precedes random variable $\max(\lambda_n^i,\rho_i\alpha_k)$ in  stop-loss order, so $X_n$ precedes $Y_n$. Moreover, the dangerous order relation is known to be a sufficient condition for the stop-loss order~\cite{hurlimann2000higher,hurlimann1998extremal,cheng1999maintenance}. Therefore, we continue our proof by showing the satisfaction of the two known conditions for dangerous order relation. In terms of the first condition, we observe that random variables $\max(\lambda_n^i,\rho_i\alpha_h)$ and $\max(\lambda_n^i,\rho_i\alpha_k)$ satisfy the once-crossing condition for crossing point $\alpha_h$. Regarding the second condition, it is simple to show that,

\begin{equation}\label{equ4}
\begin{split}
E(\max(\lambda_n^i,\rho_i\alpha_h))\leq E(\max(\lambda_n^i,\rho_i\alpha_k))
\end{split}
\end{equation}
Therefore, $\max(\lambda_n^i,\rho_i\alpha_h)$ precedes $\max(\lambda_n^i,\rho_i\alpha_k)$ in a dangerous order, and accordingly $X_n$ and $Y_n$ have the stop-order relation and the proof is complete.
\end{IEEEproof}

\subsection{Finite-size buffer shallow cloudlets}
In this section, we investigate the case when the network delay between the shallow cloudlets and the deep cloudlet is not negligible. Therefore, $\alpha=0$ is not the optimal solution since the reduction in the average loss is achieved at the expense of a higher delay.
Let $D$ be the average network delay per unit of workload (one CPU cycle) if it is served at the deep cloudlet and let's define each unit of workload as a job.
For this scenario, we enforce a deadline equal to $D$ seconds at each shallow cloudlet's buffer. In fact, a job is forwarded to the deep cloudlet only if it cannot be handled by deadline $D$.
That is, sizes of the buffers at the shallow cloudlets are calculated based on $D$ such that the maximum waiting time in each shallow cloudlet's buffer is $D$ seconds. In other words, if one unit of capacity at a shallow cloudlet can handle a job within $D$ seconds, it is not fair/justfiable to consider the allocation of that capacity to the deep cloudlet since the network delay is $D$ seconds.
Therefore, if $Q^i$ is the number of waiting jobs in the corresponding buffer of shallow cloudlet $i$ right before the arrival of a new job, the new job can be handled after $\frac{Q^i}{\rho_i\alpha}$ seconds.
If $\frac{Q^i}{\rho_i\alpha}\leq D$, then the job can be handled before the deadline $D$. Otherwise, the job is not handled before the deadline and it is forwarded to the deep cloudlet.
Therefore, we can model the deadline by a finite-size queue with length $\rho_i\alpha D$. Accordingly, the average fluid loss is calculated as
\begin{eqnarray}\label{equ5}
\overline{L}_{fb}(\alpha)=
E((\sum_{i=1}^{M}(\lambda_n^i+Q^i_{n-1}(\alpha)-\rho_i\alpha-\rho_i\alpha D)^+-(C-\alpha))^+)
\end{eqnarray}
where $Q^i_{n-1}$ is the queue length at shallow cloudlet $i$ at time $n-1$. Therefore, we propose to solve the following optimization problem,
\begin{eqnarray}
\underset{\alpha}{\text{minimize}}~\overline{L}_{fb}(\alpha)\label{equ6}
\end{eqnarray}
\begin{eqnarray}
s.t.~C_1: \sum_{i=1}^{M}E(\lambda_n^i+Q^i_{n-1}(\alpha)-\rho_i\alpha-\rho_i\alpha D)^+\leq C-\alpha\nonumber
\end{eqnarray}
\begin{eqnarray}
C_2: 0\leq\alpha\leq C\nonumber
\end{eqnarray}
where the objective is to minimize the average loss via optimizing $\alpha$ and constraint $C_1$ is required for stabilizing the queue at the deep cloudlet.

Note that the optimization problem~(\ref{equ6}) can be compared to an stop-loss reinsurance model where the objective of the problem is the stop-loss pure premium $E(X-d)^+$ with retention equal to $d=C-\alpha$~\cite{cai2007optimal,tan2009var,tan2011optimality}. Here, the retention $d=0$, i.e., a flat design with only shallow cloudlets, can be considered as the special case where the insurer transfers all loss to the reinsurer, i.e., full reinsurance. On the other hand, case $d=C$, i.e., a flat design with only a deep cloudlet, denotes the special case where the insurer retains all loss, i.e., the case that implies no reinsurance. In terms of finding the optimal solution for the reinsurance models, most of the existing studies assume that the distribution function of $X$ is known and satisfies some properties. However, here the distribution function of $X$, i.e., $\sum_{i=1}^{M}(\lambda_n^i+Q^i_{n-1}(\alpha)-\rho_i\alpha(1+D))^+$, is not known for two reasons. First, the distribution of $Q^i_{n-1}(\alpha)$ is not known. Second, even if we have the knowledge of the distribution function for $Q^i_{n-1}(\alpha)$, it is cumbersome to calculate the M-fold convolution of M pdfs. Moreover, in practice, we usually know the average of $\lambda_n^i$'s rather than their distribution function.
There are a few studies such as~\cite{hu2015optimal,reijnen2005approximations}, that consider the case when incomplete information of $X$ is available. However, those solutions are not applicable here because they either have to know at least the average and variance of $X$ or they are interested in finding the optimal retention $d$ or estimating the minimal stop-loss rather than the optimum value of $X$. Note that here we only know the average of $X$, i.e., $\sum_{i=1}^{M}E(\lambda_n^i+Q^i_{n-1}(\alpha)-\rho_i\alpha(1+D))^+$ based on the loss probability of the G/D/1 queue.
Therefore, we propose two different strategies to find the optimal value of $\alpha$. Both strategies are developed based on the Markov's inequality. That is, instead of minimizing the original objective, we minimize an upper bound calculated based on the Markov's inequality in the following theorem.
\begin{theorem}\label{Theorem2}
The objective function of optimization problem~(\ref{equ6}) is upper bounded as follows,
\begin{equation}
\overline{L}_{fb}\leq\int_{C}^{\tau}\frac{\sum_{i=1}^{M}E(\lambda_n^i+Q^i_{n-1}(\alpha)-\rho_i\alpha(1+D))^+}{C-\alpha}dx
\end{equation}
\end{theorem}
\begin{IEEEproof}
\small
\begin{eqnarray}
\begin{aligned}
&E(\sum_{i=1}^{M}(\lambda_n^i+Q^i_{n-1}(\alpha)-\rho_i\alpha-\rho_i\alpha D)^+-(C-\alpha))^+\nonumber\\
&=\int_{C}^{\infty}(x-C)dP(\sum_{i=1}^{M}(\lambda_n^i+Q^i_{n-1}(\alpha)-\rho_i\alpha-\rho_i\alpha D)^++\alpha\leq x)\nonumber\\
&=-\int_{C}^{\infty}(x-C)dP(\sum_{i=1}^{M}(\lambda_n^i+Q^i_{n-1}(\alpha)-\rho_i\alpha-\rho_i\alpha D)^++\alpha\geq x)\nonumber\\
&=\int_{C}^{\infty}P(\sum_{i=1}^{M}(\lambda_n^i+Q^i_{n-1}(\alpha)-\rho_i\alpha-\rho_i\alpha D)^++\alpha>x)dx\nonumber\\
&\approx\int_{C}^{\tau}P(\sum_{i=1}^{M}(\lambda_n^i+Q^i_{n-1}(\alpha)-\rho_i\alpha-\rho_i\alpha D)^++\alpha>x)dx\nonumber\\
\end{aligned}
\end{eqnarray}
\normalsize
where $\tau$ in the approximation can be decided based on the tail of the distribution of $\lambda_n^i$ such that $P(\sum_{i=1}^{M}(\lambda_n^i+Q^i_{n-1}(\alpha)-\rho_i\alpha-\rho_i\alpha D)^++\alpha>\tau)\leq\epsilon$, i.e., 
 \begin{eqnarray}
 \begin{aligned}
 &\int_{\tau}^{\infty}P(\sum_{i=1}^{M}(\lambda_n^i+Q^i_{n-1}(\alpha)-\rho_i\alpha-\rho_i\alpha D)^++\alpha>x)dx\nonumber\\
 &<<\int_{C}^{\tau}P(\sum_{i=1}^{M}(\lambda_n^i+Q^i_{n-1}(\alpha)-\rho_i\alpha-\rho_i\alpha D)^++\alpha>x)dx
\end{aligned}
\end{eqnarray}
Then, we have
\begin{eqnarray}
\begin{aligned}
&\int_{C}^{\tau}P(\sum_{i=1}^{M}(\lambda_n^i+Q^i_{n-1}(\alpha)-\rho_i\alpha-\rho_i\alpha D)^++\alpha>x)dx\nonumber\\
&\leq\int_{C}^{\tau}P(\sum_{i=1}^{M}(\lambda_n^i+Q^i_{n-1}(\alpha)-\rho_i\alpha-\rho_i\alpha D)^+>C-\alpha)dx\nonumber\\
&\leq\int_{C}^{\tau}\frac{\sum_{i=1}^{M}E(\lambda_n^i+Q^i_{n-1}(\alpha)-\rho_i\alpha(1+D))^+}{C-\alpha}dx\nonumber~~~~~~~~~~~~~~~~~
\end{aligned}
\end{eqnarray}
where the last inequality is in accordance with the Markov's inequality. The proof is complete.
\end{IEEEproof}

\subsubsection{G/D/1 loss probability approach}
In the first approach, we rely on the loss probability of the G/D/1 queue.
According to queueing analysis~\cite{kim2001loss}, we have,
\begin{equation}
E(\lambda_n^i+Q^i_{n-1}(\alpha)-\rho_i\alpha(1+D))^+=P_i(\alpha)\overline{\lambda}_i
\end{equation}
where $P_i(\alpha)$ is the loss probability of the finite-size queue and can be accurately estimated from the tail
probability (overflow probability) of an infinite buffer system as follows~\cite{kim2001loss},
\begin{eqnarray}\label{equ55}
P_{i}(\alpha)=\gamma_i(\alpha) e^{-\frac{1}{2}\underset{n\geq1}{\min}M_{n}^i(\alpha)},
\end{eqnarray}
where
\begin{eqnarray}
\gamma_i(\alpha)=\frac{1}{\overline{\lambda}_i\sqrt{2\pi}\sigma_i}e^\frac{(\rho_i\alpha-\overline{\lambda}_i)^2}{2\sigma_{i}^2}\int_{\rho_i\alpha}^{\infty}
(r-\rho_i\alpha)e^\frac{-(r-\overline{\lambda}_i)^2}{2\sigma_i^2}dr,&
\end{eqnarray}
and for each $n\geq1$,
\begin{eqnarray}
M_{n}^i(\alpha)=\frac{(\rho_i\alpha D+n(\rho_i\alpha-\overline{\lambda}_i))^2}
{nC_{\lambda_n^i}(0)+2\sum_{l=1}^{n-1}(n-l)C_{\lambda_n^i}(l)},
\end{eqnarray}
and $C_{\lambda_n^i}(l)$ is the autocovariance of $\lambda_n^i$ probability function and we have $\sigma^{2}_{i}=C_{\lambda_n^i}(0)$.
Note that function~(\ref{equ55}) is valid when $\rho_i\alpha\geq\overline{\lambda}_i$, i.e., when $\alpha\geq\sum_{i=1}^{M}\overline{\lambda}_i$. In addition, it is known that the estimation yields the highest level of accuracy when $\lambda_n^i$ is characterized by a Gaussian process. Therefore, in this approach, we focus on the case that the input process to each queue, i.e., $\lambda_n^i$, follows a Gaussian process and propose to solve the following optimization problem,
\begin{eqnarray}\label{equ7}
\underset{\alpha}{\text{minimize}}~\frac{\sum_{i=1}^{M}P_i(\alpha)\overline{\lambda}_i}{C-\alpha}
\end{eqnarray}
\begin{eqnarray}
s.t.~C_1: \sum_{i=1}^{M}P_i(\alpha)\overline{\lambda}_i\leq C-\alpha\nonumber
\end{eqnarray}
\begin{eqnarray}
C_2: \sum_{i=1}^{M}\overline{\lambda}_i\leq\alpha< C\nonumber
\end{eqnarray}
\begin{algorithm}
\caption{}
\label{alg:3}
\begin{algorithmic}[1]
\STATE {find a feasible stepsize $\epsilon\geq0$}
\STATE {$r\gets1+\epsilon$}
\STATE {$\hat{\alpha}\gets C$}
\REPEAT
\STATE{solve problem~(\ref{equ8}) for $\alpha$ in range~(\ref{range}) and find $\accentset{\star}{\alpha}$}
\IF{$\accentset{\star}{\alpha}\neq\emptyset$}
\STATE{$\hat{\alpha}\gets\accentset{\star}{\alpha}$}
\STATE{$r\gets\frac{C-\hat{\alpha}}{\sum_{i=1}^{M}P_i(\hat{\alpha})\overline{\lambda}_i}+\epsilon$}
\ENDIF
\UNTIL{$\accentset{\star}{\alpha}=\emptyset$}
\end{algorithmic}
\end{algorithm}

To solve optimization problem~(\ref{equ7}), we propose a centralized heuristic algorithm. Our algorithm is motivated by two observations. First, $p_i(\alpha)$ is a non-increasing function with respect to $\alpha$ when $\alpha\geq\sum_{i=1}^{M}\overline{\lambda}_i$ ~\cite{ghamkhari2013energy,kiani2015profit}. Second, alternative optimization problem~(\ref{equ8}) is a convex optimization problem if $\alpha$ is limited to some specific range and can be solved efficiently by interior point methods.
In other words, problem~(\ref{equ7}) is generally nonconvex. Therefore, we introduce a new variable $r$ such that $r=\frac{C-\alpha}{\sum_{i=1}^{M}P_i(\alpha)\overline{\lambda}_i}$. Accordingly, inspired by coordinate descent techniques~\cite{bezdek1987local}, we solve successively alternate minimizations~(\ref{equ8}) in $\alpha$ while holding $r$ fixed.
As shown in~ Algorithm~\ref{alg:3}, we first choose a feasible value for stepsize $\epsilon$. Note that Algorithm~\ref{alg:3} converges to the optimal solution provided that the stepsize is selected small enough. We also set initial ratio $r=1+\epsilon$ and $C$ is chosen as the initial solution. Then, we solve the following optimization problem for the given value of $r$,
\begin{eqnarray}\label{equ8}
\underset{\alpha}{\text{minimize}}~\sum_{i=1}^{M}P_i(\alpha)\overline{\lambda}_i
\end{eqnarray}
\begin{eqnarray}
s.t.~C_1: \sum_{i=1}^{M}P_i(\alpha)\overline{\lambda}_i-\frac{C-\alpha}{r}\leq 0\nonumber
\end{eqnarray}

\begin{eqnarray}
C_2: \sum_{i=1}^{M}\overline{\lambda}_i\leq\alpha< C\nonumber
\end{eqnarray}
Finally, we update the ratio $r$ and optimal solution $\hat{\alpha}$ as shown in Algorithm~\ref{alg:3}. We repeat this procedure until there is no optimal solution for problem~(\ref{equ8}). The convexity of problem~(\ref{equ8}) is proven in the following theorem.
\begin{theorem}\label{Theorem3}
The constrained optimization problem~(\ref{equ8}) is a
convex optimization problem if $\alpha$ is limited to,
\begin{eqnarray}\label{range}
\alpha\in\sum_{i=1}^M\overline{\lambda}_i+[\underset{i}{\max}~.07071\frac{\sigma_i}{\rho_i},\underset{i}{\min}~1.4477\frac{\sigma_i}{\rho_i}]
\end{eqnarray}
\end{theorem}

\begin{IEEEproof}
To show the convexity of the proposed optimization problem, we are required to prove~\cite{boyd2004convex}:
\begin{itemize}
\item{The objective function, i.e., $\sum_{i=1}^{M}P_i(\alpha)\overline{\lambda}_i$, is convex.}
\item{The inequality constraint $C_1$ is convex.}
\end{itemize}
We start by proving the convexity of $P_i(\alpha)$, i.e., loss probability function. It is known that the loss probability is a convex function when the service rate $\rho_i\alpha$~\cite{ghamkhari2013energy,kiani2015profit} is limited to,
\begin{eqnarray}
\rho_i\alpha\in[\overline{\lambda}_i+.07071\sigma_i, \overline{\lambda}_i+1.4477\sigma_i]
\end{eqnarray}
Accordingly, $P_i(\alpha)$ is a convex function for all $i$ if,
\begin{eqnarray}
\alpha\in\sum_{i=1}^M\overline{\lambda}_i+[\underset{i}{\max}~.07071\frac{\sigma_i}{\rho_i},\underset{i}{\min}~1.4477\frac{\sigma_i}{\rho_i}]
\end{eqnarray}
Then, the inequality constraint function of $C_1$ and the objective function are both proven to be convex since they are summations of convex functions, and the proof is complete.
\end{IEEEproof}
An interesting extension for the optimization problem~(\ref{equ7}) is
the case when the loss probability at each shallow cloudlet $i$ is upper bounded by a constant $TH_i$. In other words, this extension limits the number of jobs that can be forwarded to the deep cloudlet from the shallow cloudlets. Therefore, we incorporate this requirement
into our optimization problem by adding the inequality constraints $P_i(\alpha)<TH_i$ as follows,
\begin{eqnarray}\label{equ10}
\underset{\alpha}{\text{minimize}}~\frac{\sum_{i=1}^{M}P_i(\alpha)\overline{\lambda}_i}{C-\alpha}
\end{eqnarray}
\begin{eqnarray}
s.t.~C_1: \sum_{i=1}^{M}P_i(\alpha)\overline{\lambda}_i\leq C-\alpha\nonumber
\end{eqnarray}
\begin{eqnarray}
C_2: P_{i}(\alpha)\leq TH_i~\forall i=1,...,M\nonumber
\end{eqnarray}
\begin{eqnarray}
C_3: \sum_{i=1}^{M}\overline{\lambda}_i\leq\alpha< C\nonumber
\end{eqnarray}
Note that the new inequality constraints $C_2$ form a convex set under the same requirement as Theorem~\ref{Theorem3}. Therefore, Algorithm~\ref{alg:3} can still be used to solve problem~(\ref{equ10}).
\subsubsection{Queue length estimation approach}
In the previous approach, we rely on the accuracy of loss probability of a G/D/1 queue and replace loss $E(\lambda_n^i+Q^i_{n-1}(\alpha)-\rho_i\alpha(1+D))^+$ with $P_i(\alpha)\overline{\lambda}_i$. However, as mentioned earlier, function $P_i(\alpha)$ is accurate when the input process $\lambda_n^i$ is characterized by a Gaussian distribution, and more importantly, it is derived based on the assumption that $\rho_i\alpha\geq\overline{\lambda}_i$. Therefore, in this section, we propose another approach which can be accurate for other distributions such as the uniform distribution and is valid for all values of $\alpha$. The idea is to replace the queue length $Q^i_{n-1}$ in $E(\lambda_n^i+Q^i_{n-1}(\alpha)-\rho_i\alpha(1+D))^+$ with a linear estimation of the Average Queue Length (AQL). We propose the following linear estimation,
\begin{eqnarray}
e_{AQL_i}=\
\scalebox{1}{$
\begin{cases}
0,&\overline{\lambda}_i\leq\rho_i\alpha\\
\\
a\alpha+b,&\rho_i\alpha<\overline{\lambda}_i\leq\rho_i\alpha(1+D)\\
\\
\rho_i\alpha D,&\overline{\lambda}_i>\rho_i\alpha(1+D)\\
\end{cases}$}
\end{eqnarray}
where constants $a$ and $b$ can be calculated by solving two equations $a(\frac{\sum_{i=1}^M\overline{\lambda}_i}{1+D})+b=\rho_i D(\frac{\sum_{i=1}^M\overline{\lambda}_i}{1+D})$ and $a(\sum_{i=1}^M\overline{\lambda}_i)+b=0$. After reordering, we have
\begin{eqnarray}\label{equ11}
e_{AQL_i}=\
\scalebox{1}{$
\begin{cases}
0,&\alpha\geq\sum_{i=1}^M\overline{\lambda}_i\\
\\
-\rho_i\alpha+\rho_i\sum_{i=1}^M\overline{\lambda}_i,&\frac{\sum_{i=1}^M\overline{\lambda}_i}{1+D}\leq\alpha<\sum_{i=1}^M\overline{\lambda}_i\\
\\
\rho_i\alpha D,&\alpha<\frac{\sum_{i=1}^M\overline{\lambda}_i}{1+D}\\
\end{cases}$}
\end{eqnarray}
Note that estimation~(\ref{equ11}) yields a higher accuracy for a smaller variance of $\lambda_n^i$. In case that the variance is not small, we can adjust the estimation as follows
\begin{eqnarray}
e_{AQL_i}=~~~~~~~~~~~~~~~~~~~~~~~~~~~~~~~~~~~~~~~~~~~~~~~\nonumber
\end{eqnarray}
\vspace{-.2in}
\begin{eqnarray}
\scalebox{.9}{$
\begin{cases}
0,&\alpha\geq\sum_{i=1}^M\overline{\lambda}_i+\kappa_i\\
\\
-\rho_i(\alpha-\kappa_i)+\rho_i\sum_{i=1}^M\overline{\lambda}_i,&\frac{\sum_{i=1}^M\overline{\lambda}_i}{1+D}+\kappa_i\leq\alpha<\sum_{i=1}^M\overline{\lambda}_i+\kappa_i\\
\\
\rho_i(\alpha-\kappa_i) D,&\alpha<\frac{\sum_{i=1}^M\overline{\lambda}_i}{1+D}+\kappa_i\\
\end{cases}$}
\end{eqnarray}
\begin{figure*}[htb]
        \begin{subfigure}{0.33\textwidth}
                \includegraphics[width=1\linewidth]{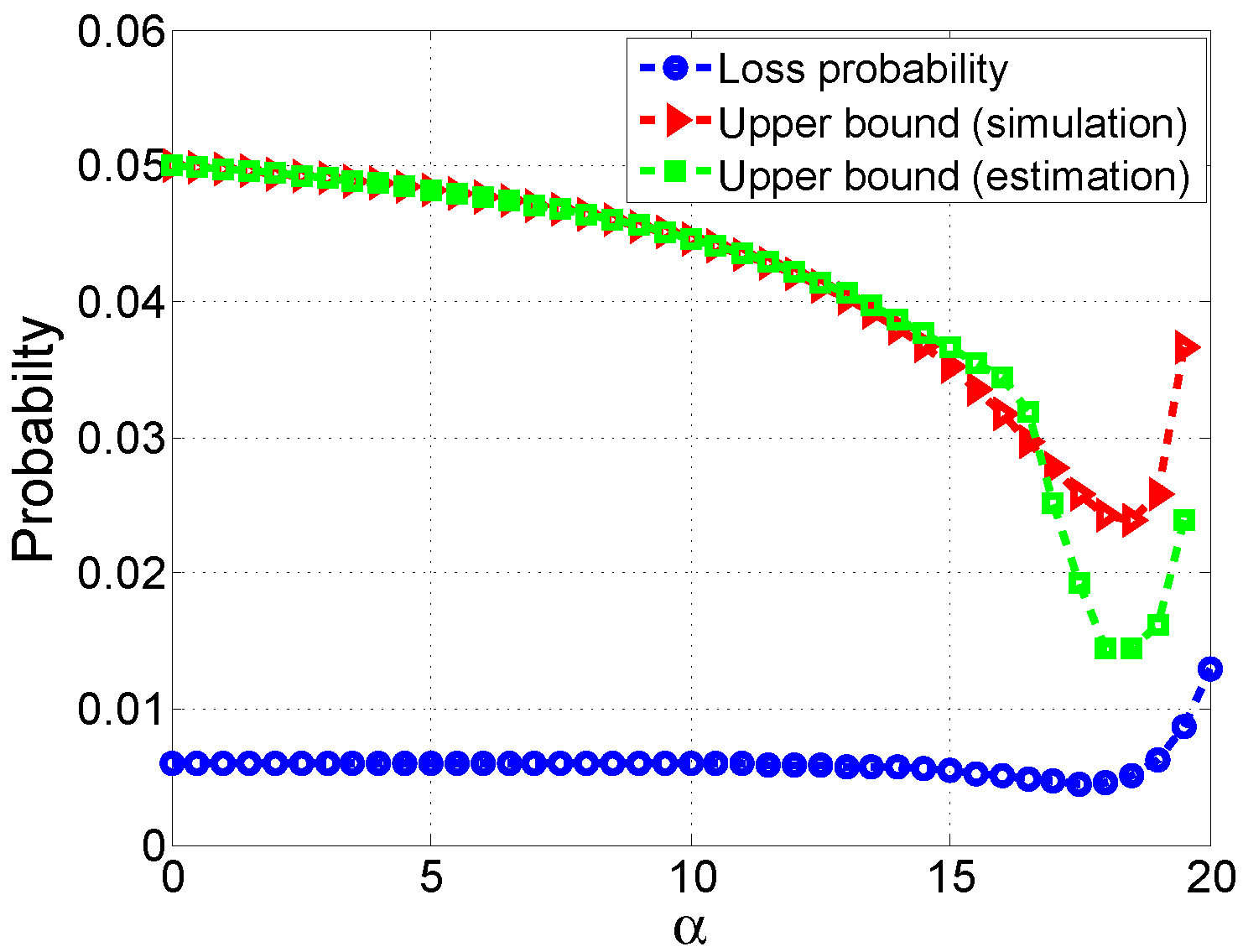}
                \caption{When the input is a Gaussian AR process.}
                \label{fig:21}
        \end{subfigure}
        \begin{subfigure}{0.33\textwidth}
                \includegraphics[width=1.05\linewidth]{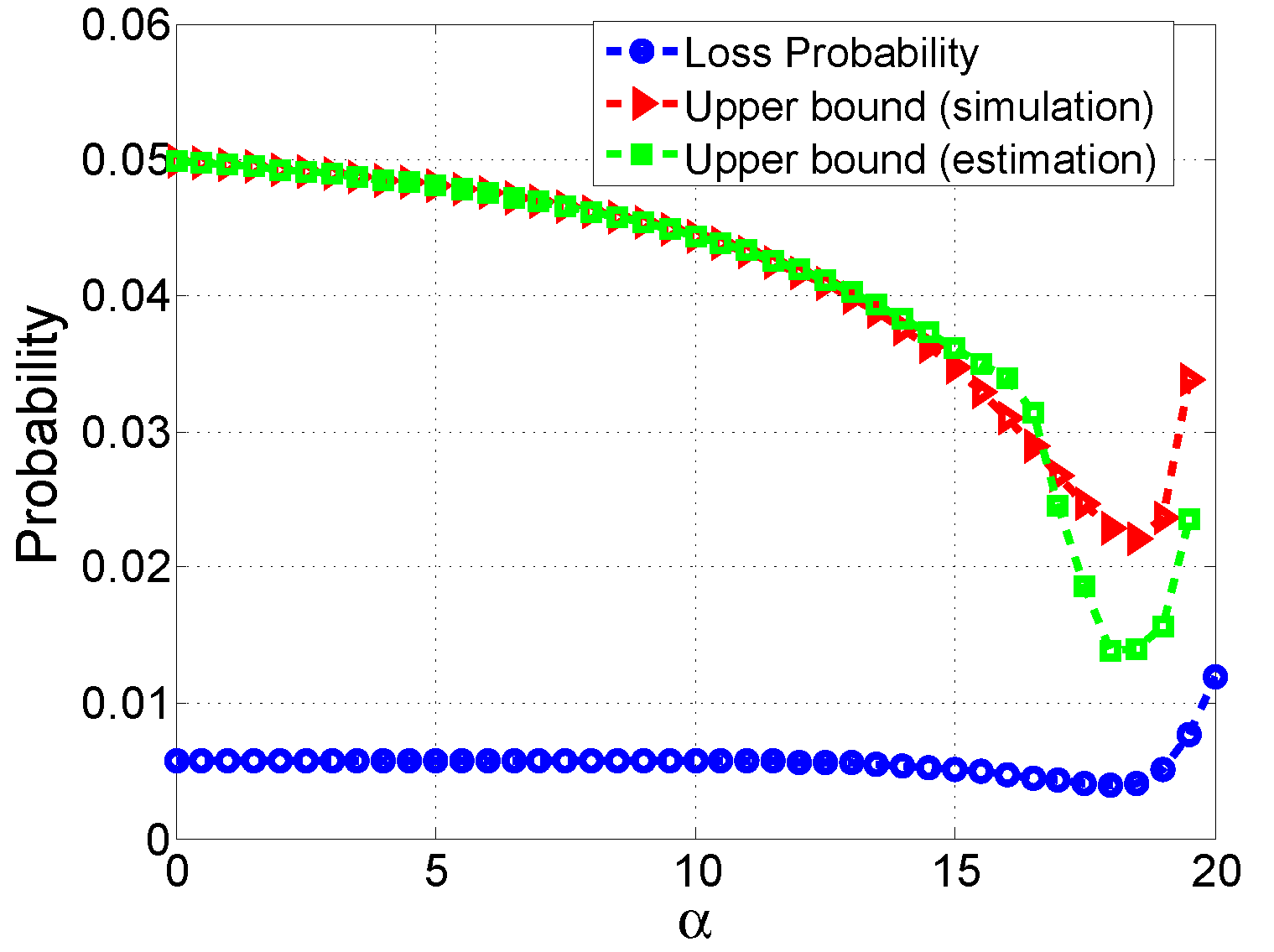}
                \caption{When the input is a Gaussian process.}
                \label{fig:22}
        \end{subfigure}
        \begin{subfigure}{0.33\textwidth}
                \includegraphics[width=1.05\linewidth]{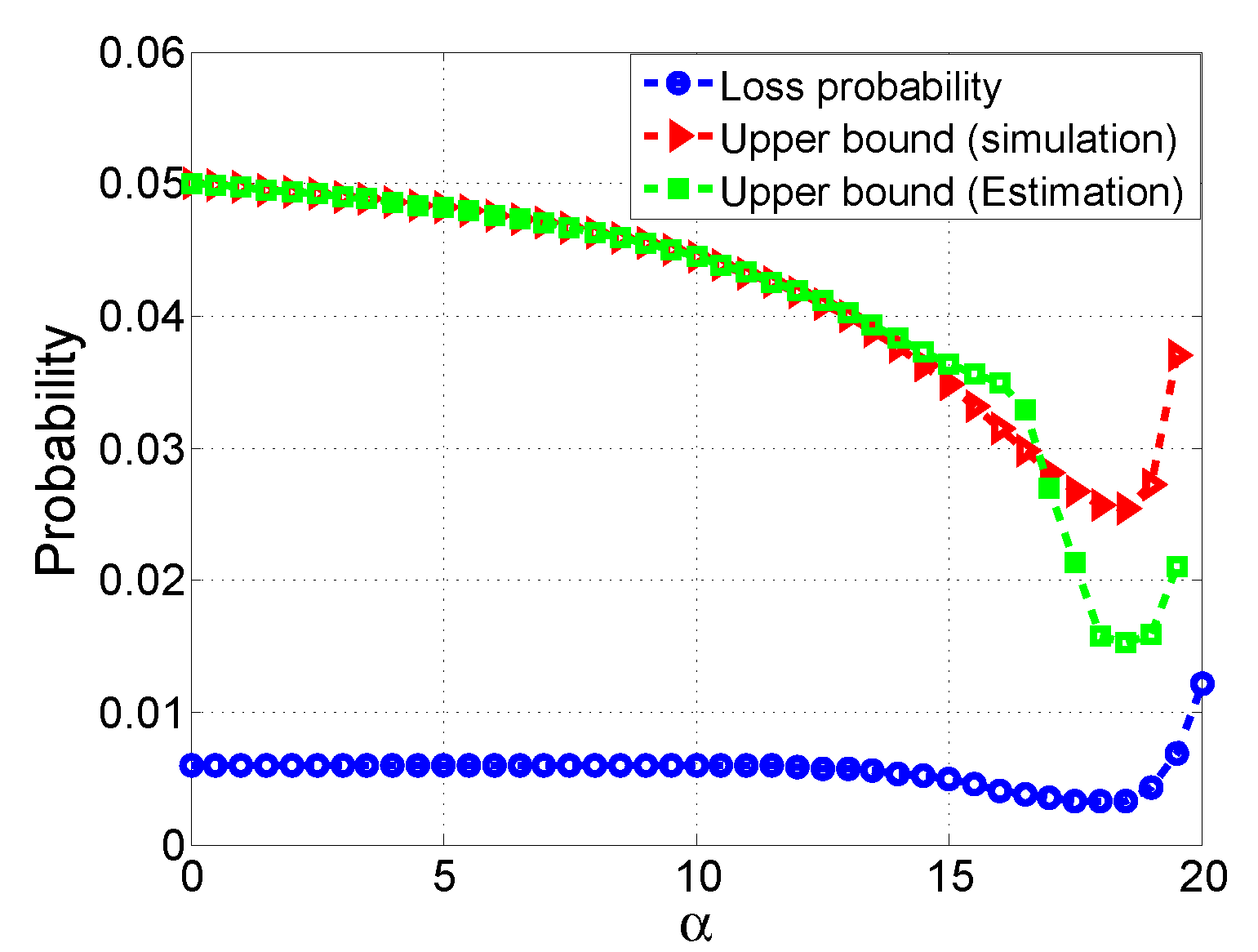}
                \caption{When the input is a uniform process.}
                \label{fig:23}
        \end{subfigure}%
                \caption{The comparison between the shape of the loss probability with the shape of the proposed upper bound versus $\alpha$ for $D=0.1$.}\label{fig:2}
\end{figure*}
\begin{figure}
\epsfig{file=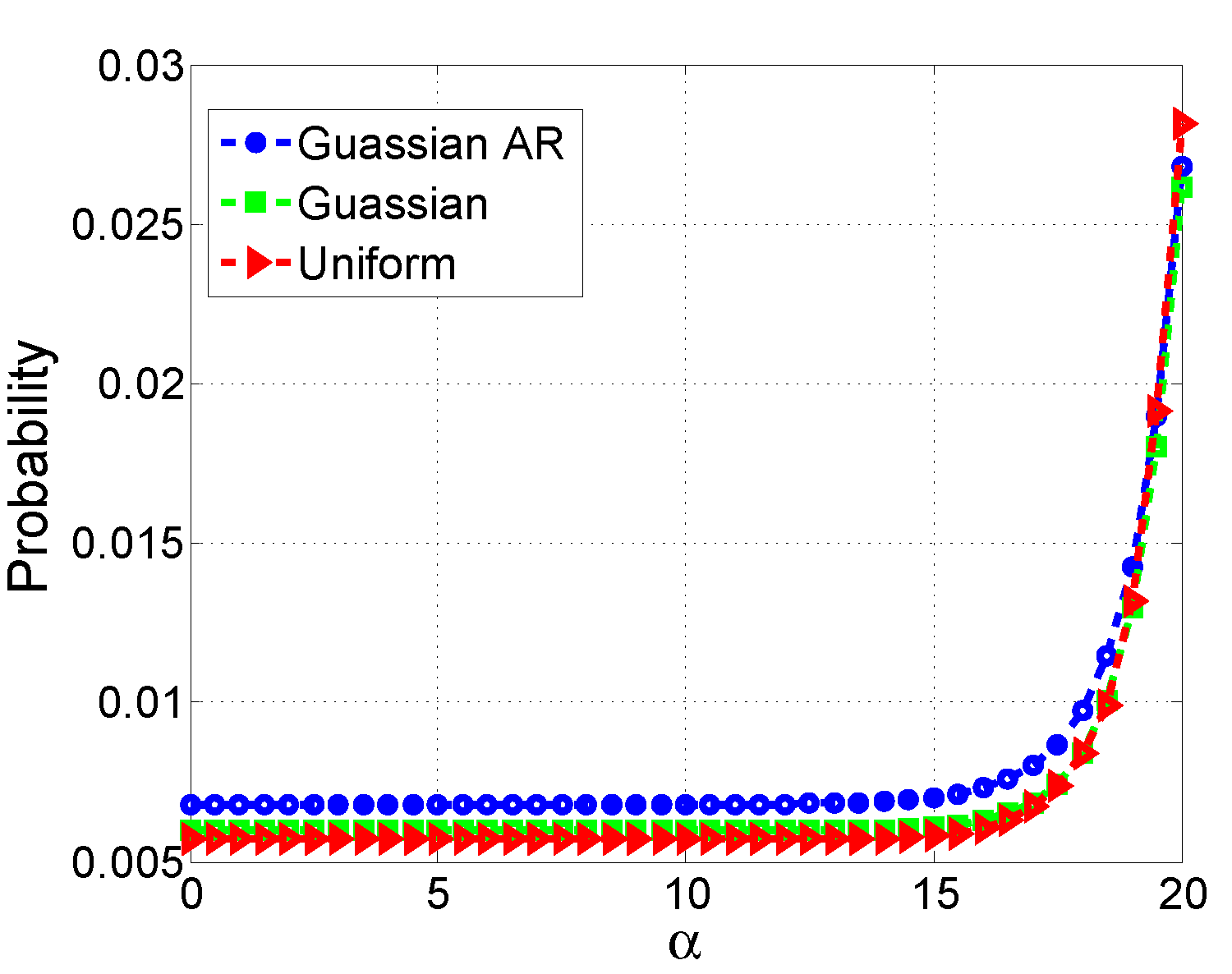,width=.8\linewidth,clip=}
\caption{Loss probability versus $\alpha$ for different input processes and when $D=0$.}
\label{fig:3}
\end{figure}
where constant $\kappa_i$ is calculated heuristically and according to the variance of $\lambda_n^i$.
Therefore, in order to find an approximate solution, we can replace the optimization problem~(\ref{equ7}) with the following problem,
\begin{eqnarray}\label{equ12}
\underset{\alpha}{\text{minimize}}~\frac{\sum_{i=1}^{M}E(\lambda_n^i+e_{AQL_i}-\rho_i\alpha(1+D))^+}{C-\alpha}
\end{eqnarray}
\begin{eqnarray}
s.t.~C_1: \sum_{i=1}^{M}E(\lambda_n^i+e_{AQL_i}-\rho_i\alpha(1+D))^+\leq C-\alpha\nonumber
\end{eqnarray}
\begin{eqnarray}
C_2: 0\leq\alpha\leq C\nonumber
\end{eqnarray}
The same procedure as Algorithm~\ref{alg:3} is still valid to solve problem~(\ref{equ12}) for two reasons. That is, function $E(\lambda_n^i+e_{AQL_i}-\rho_i\alpha(1+D))^+$ is a non-increasing and convex function with respect to $\alpha$ as proved in the following theorem.
\begin{theorem}
Function $g_i(\alpha):=E(\lambda_n^i+e_{AQL_i}-\rho_i\alpha(1+D))^+$ is a non-increasing and convex function with respect to $\alpha$.
\end{theorem}
\begin{IEEEproof}
\begin{eqnarray}
g_i(\alpha)=\int_{\rho_i\alpha(1+D)-e_{AQL_i}}^{\infty}(x-\rho_i\alpha(1+D)+e_{AQL_i})
f_{\lambda_n^i}(x)dx\nonumber
\end{eqnarray}
\begin{eqnarray}
\end{eqnarray}
Then, according to Leibniz integral rule, we have
\begin{eqnarray}
g'_i(\alpha)=
\int_{\rho_i\alpha(1+D)-e_{AQL_i}}^{\infty}(-\rho_i(1+D)+e'_{AQL_i})
f_{\lambda_n^i}(x)dx\nonumber
\end{eqnarray}
\begin{eqnarray}
\end{eqnarray}

where $-\rho_i(1+D)+e'_{AQL_i}\leq-\rho_i$ and thus $g'_i(\alpha)\leq0$.
Therefore, function $g_i(\alpha)$ is proven to be non-increasing.
Moreover, by taking the second derivative with respect to $\alpha$, we have
\begin{eqnarray}
g_i''(\alpha)=
(\rho_i(1+D)-e'_{AQL_i})^2f_{\lambda_n^i}(\rho_i\alpha(1+D)-e_{AQL_i})\geq0\nonumber
\end{eqnarray}
\begin{eqnarray}
\end{eqnarray}
Therefore, $g_i(\alpha)$ is convex and the proof is complete.
\end{IEEEproof}

\section{Simulation Results}\label{sec:simulations}

In this section, we evaluate the performance of the proposed upper bound for the average loss based on both randomly generated input and real trace data. In both cases, we consider a fog computing network consisting of three shallow cloudlets connected to a deep cloudlet, i.e., a network architecture similar to Fig.~\ref{fig:1}.
\begin{figure*}
        \begin{subfigure}{0.33\textwidth}
                \includegraphics[width=1\linewidth]{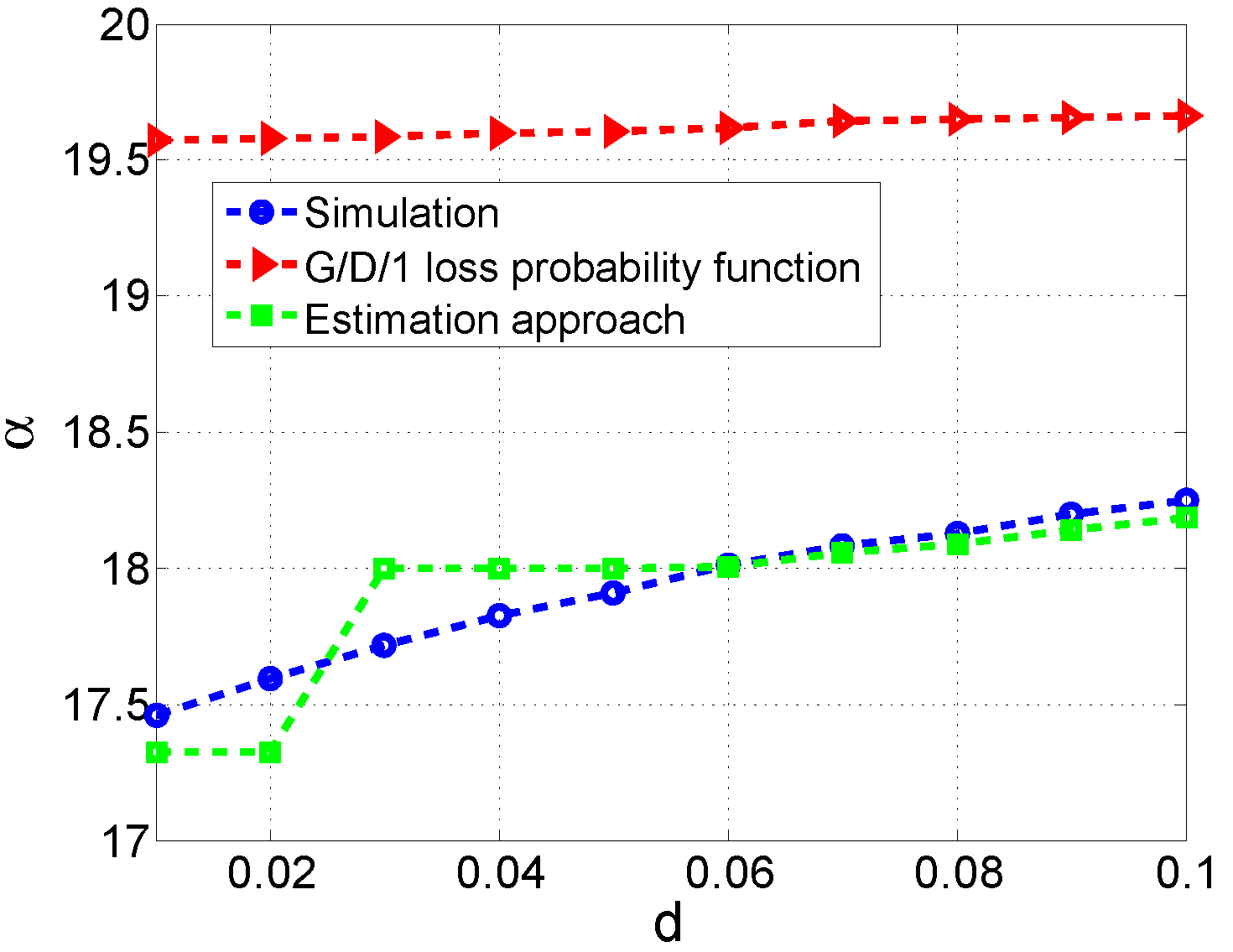}
                \caption{When the input is a Gaussian AR process.}
                \label{fig:41}
        \end{subfigure}
        \begin{subfigure}{0.33\textwidth}
                \includegraphics[width=1.05\linewidth]{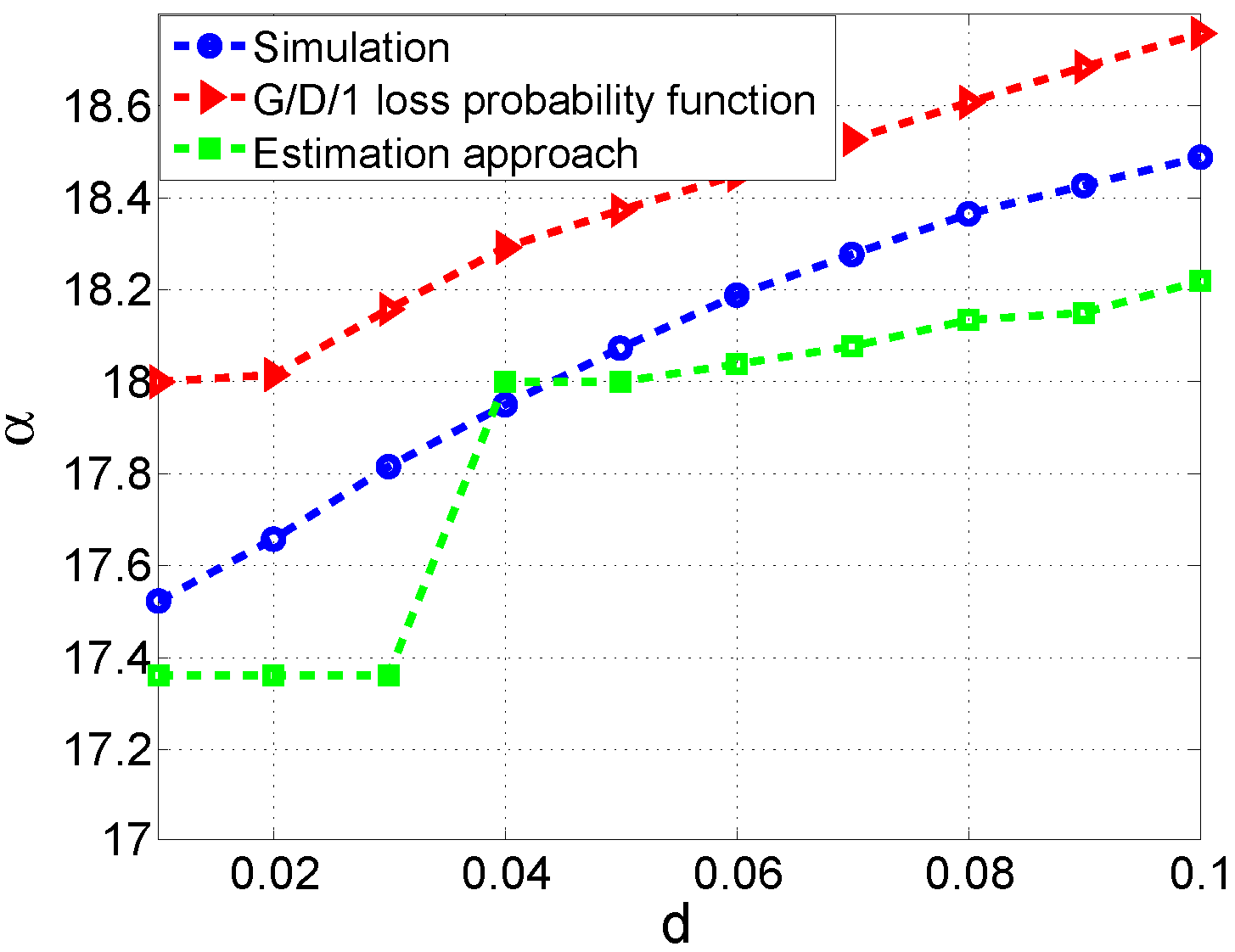}
                \caption{When the input is a Gaussian process.}
                \label{fig:42}
        \end{subfigure}
        \begin{subfigure}{0.33\textwidth}
                \includegraphics[width=1.05\linewidth]{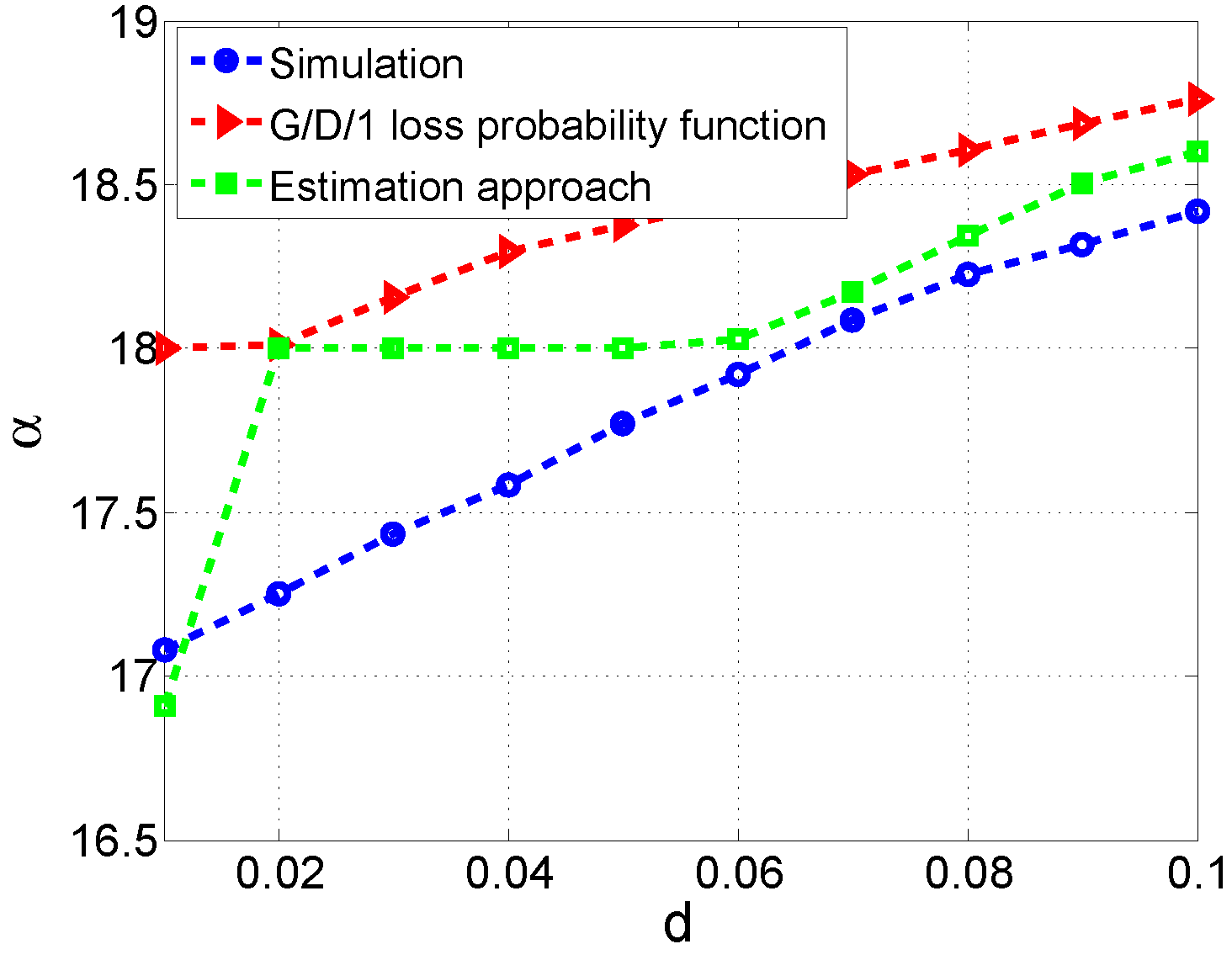}
                \caption{When the input is a uniform process.}
                \label{fig:43}
        \end{subfigure}%
                \caption{Optimal $\alpha$ versus $D$.}\label{fig:4}
\end{figure*}

\begin{figure*}
        \begin{subfigure}{0.33\textwidth}
                \includegraphics[width=1\linewidth]{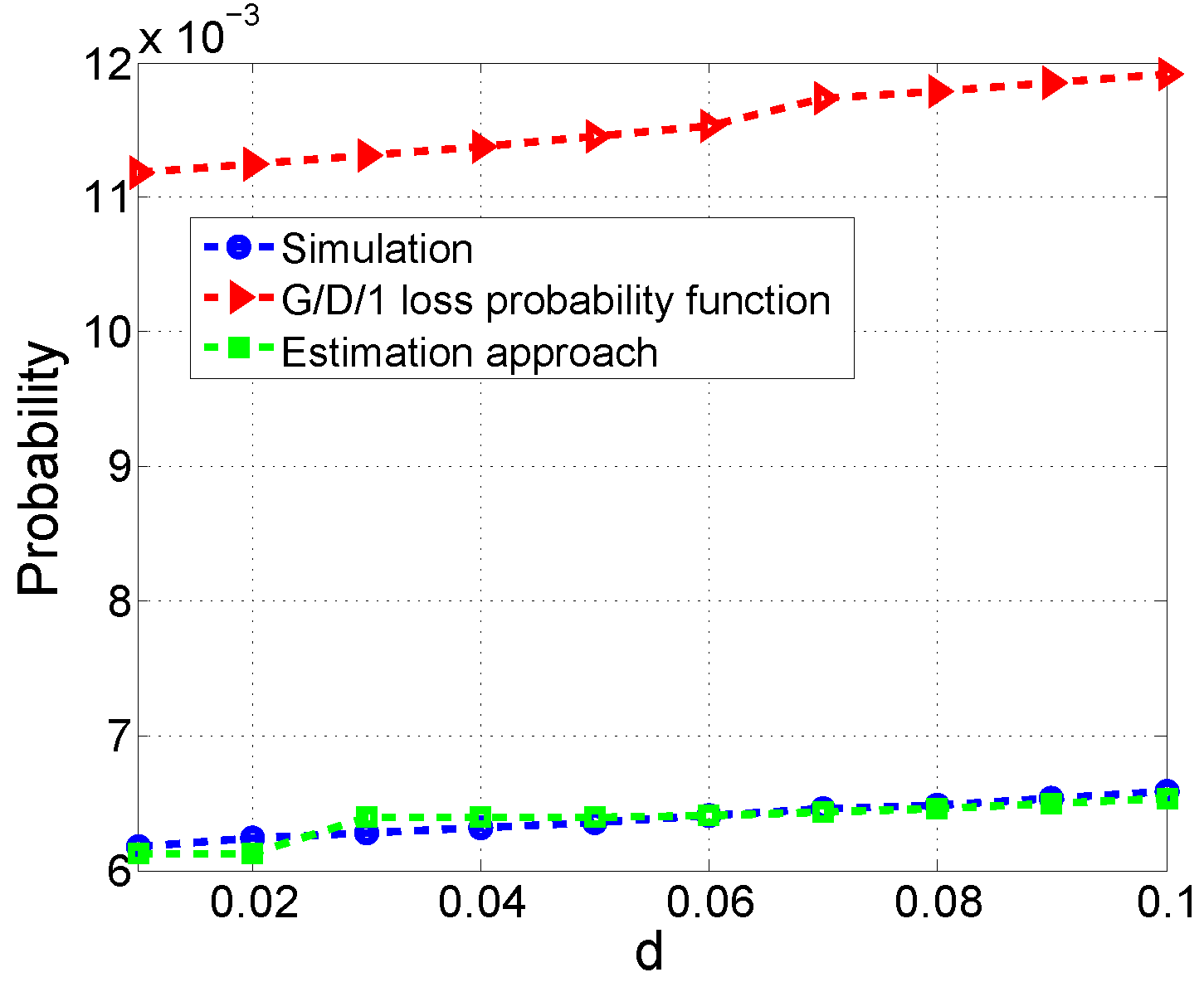}
                \caption{When the input is a Gaussian AR process.}
                \label{fig:51}
        \end{subfigure}
        \begin{subfigure}{0.33\textwidth}
                \includegraphics[width=1.05\linewidth]{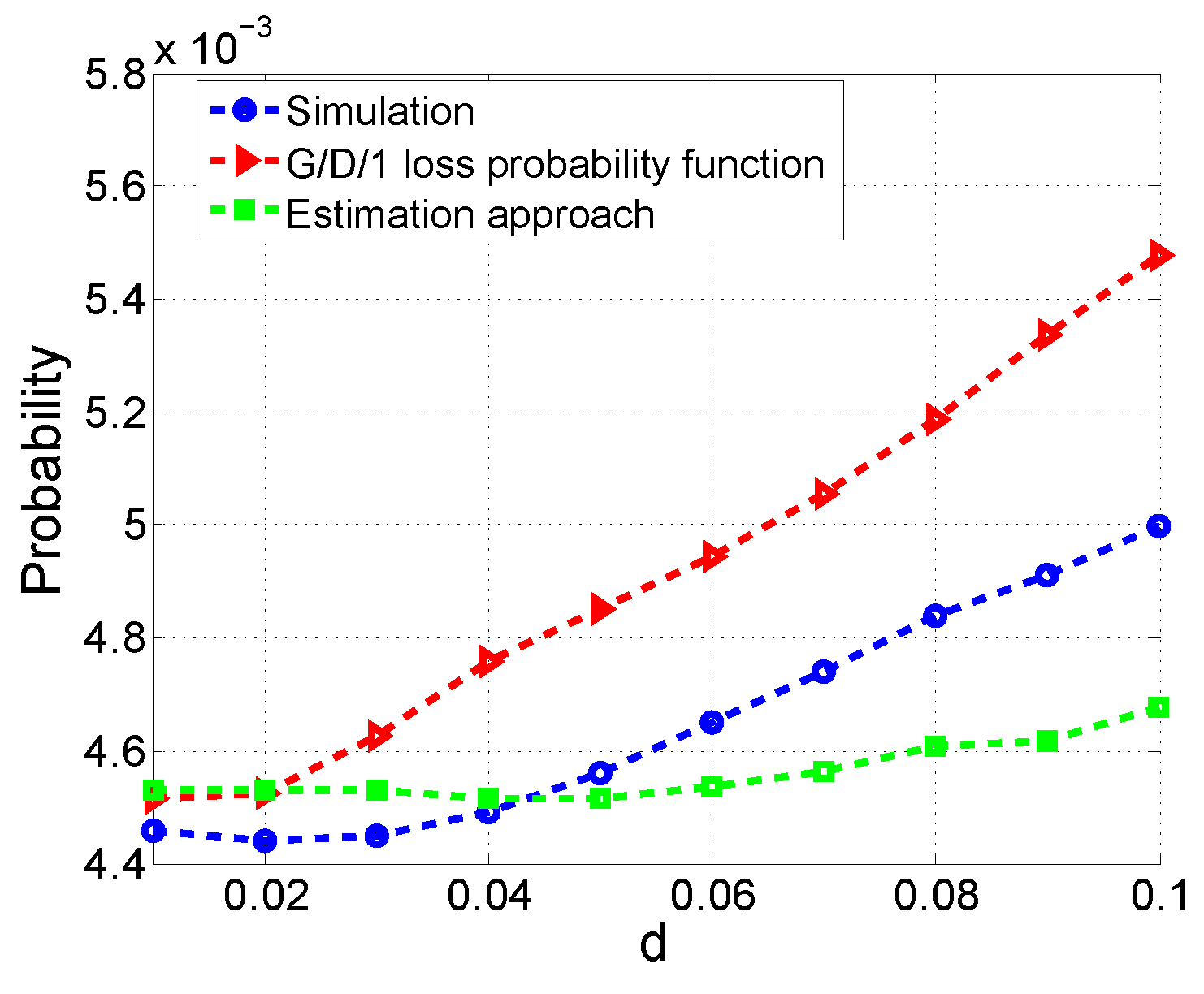}
                \caption{When the input is a Gaussian process.}
                \label{fig:52}
        \end{subfigure}
        \begin{subfigure}{0.33\textwidth}
                \includegraphics[width=1.05\linewidth]{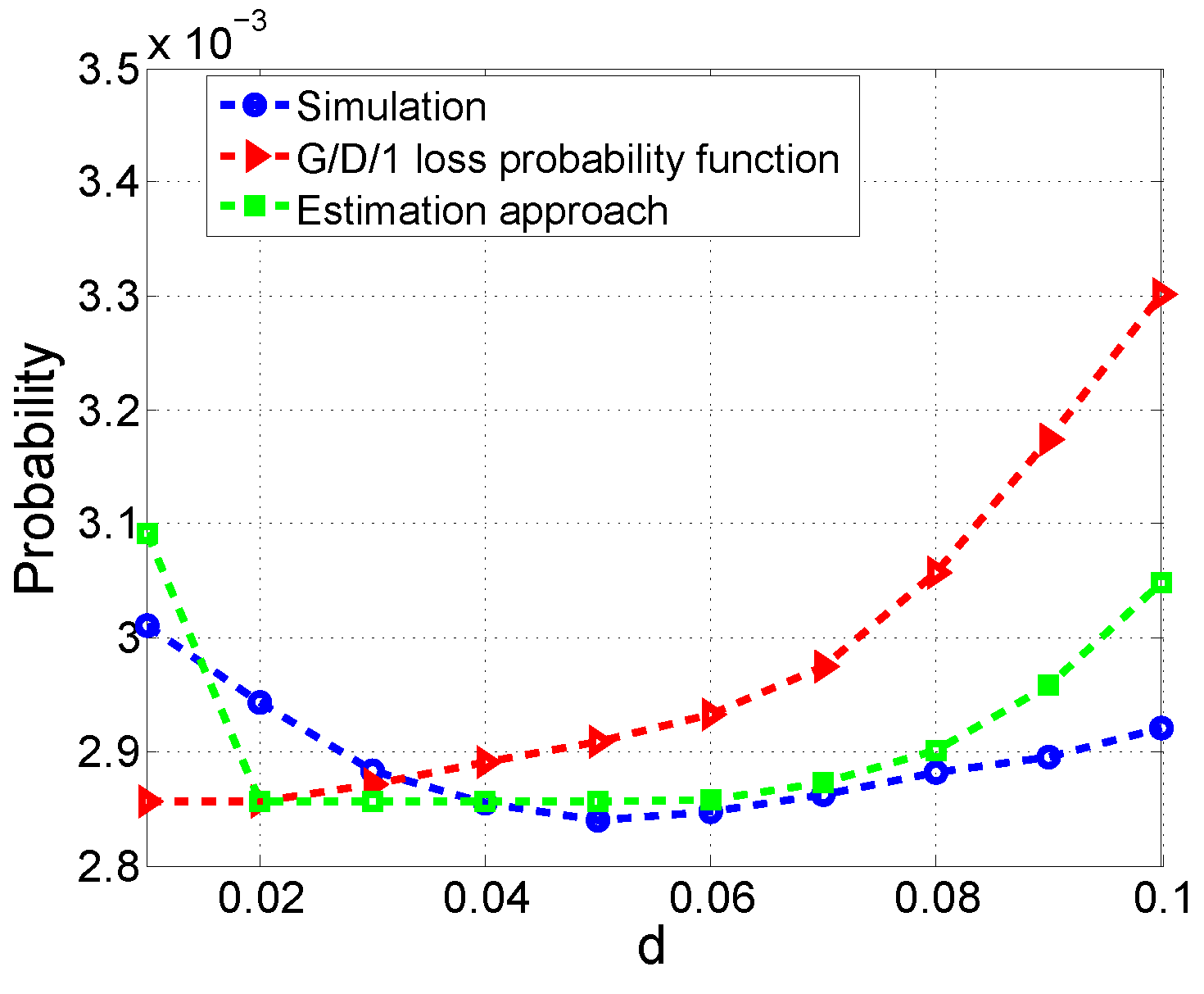}
                \caption{When the input is a uniform process.}
                \label{fig:53}
        \end{subfigure}%
                \caption{Optimum loss probability versus $D$.}\label{fig:5}
\end{figure*}

\begin{figure*}
        \begin{subfigure}{0.33\textwidth}
                \includegraphics[width=1\linewidth]{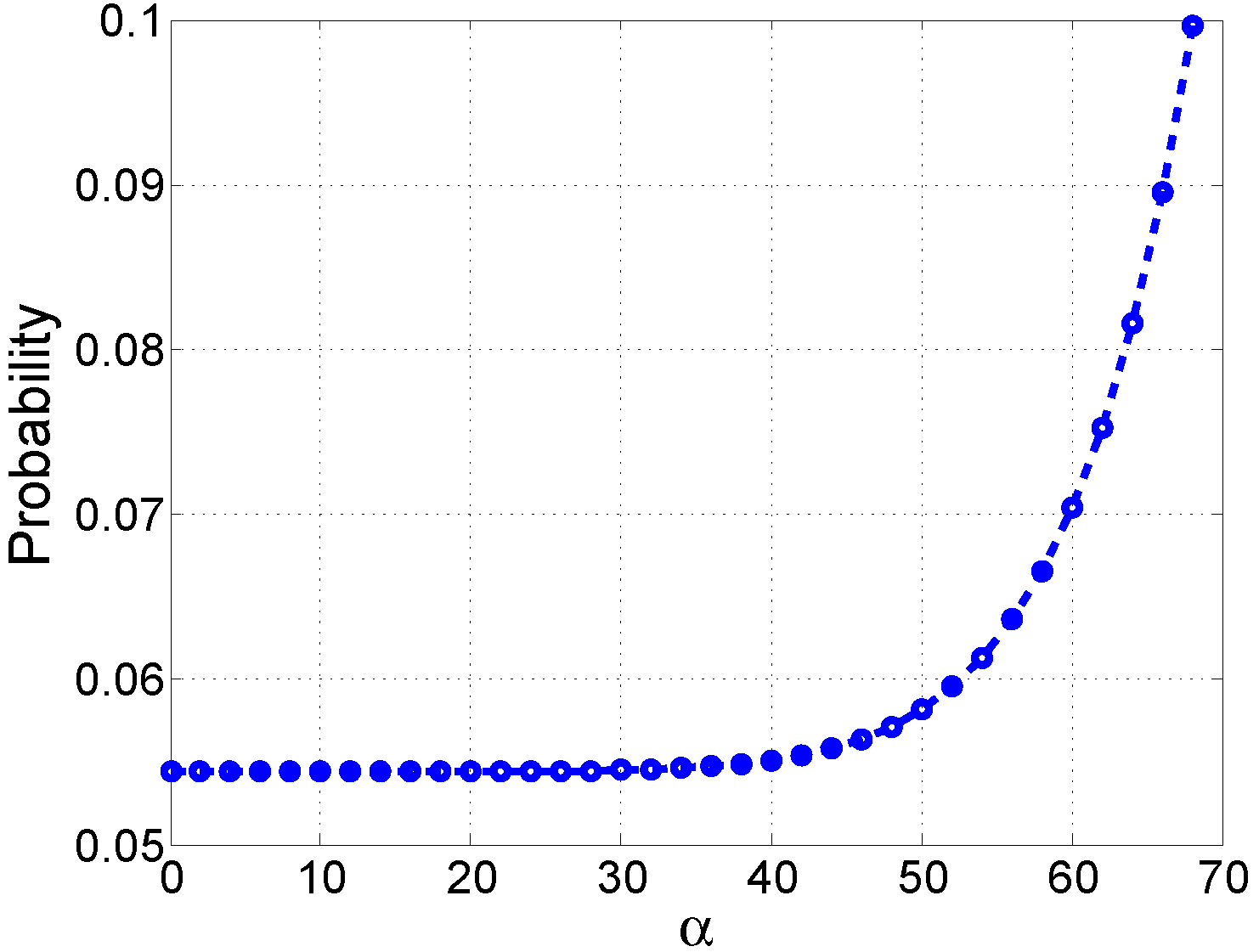}
                \caption{Loss probability versus $\alpha$ when $D=0$.}
                \label{fig:61}
        \end{subfigure}
        \begin{subfigure}{0.33\textwidth}
                \includegraphics[width=1.05\linewidth]{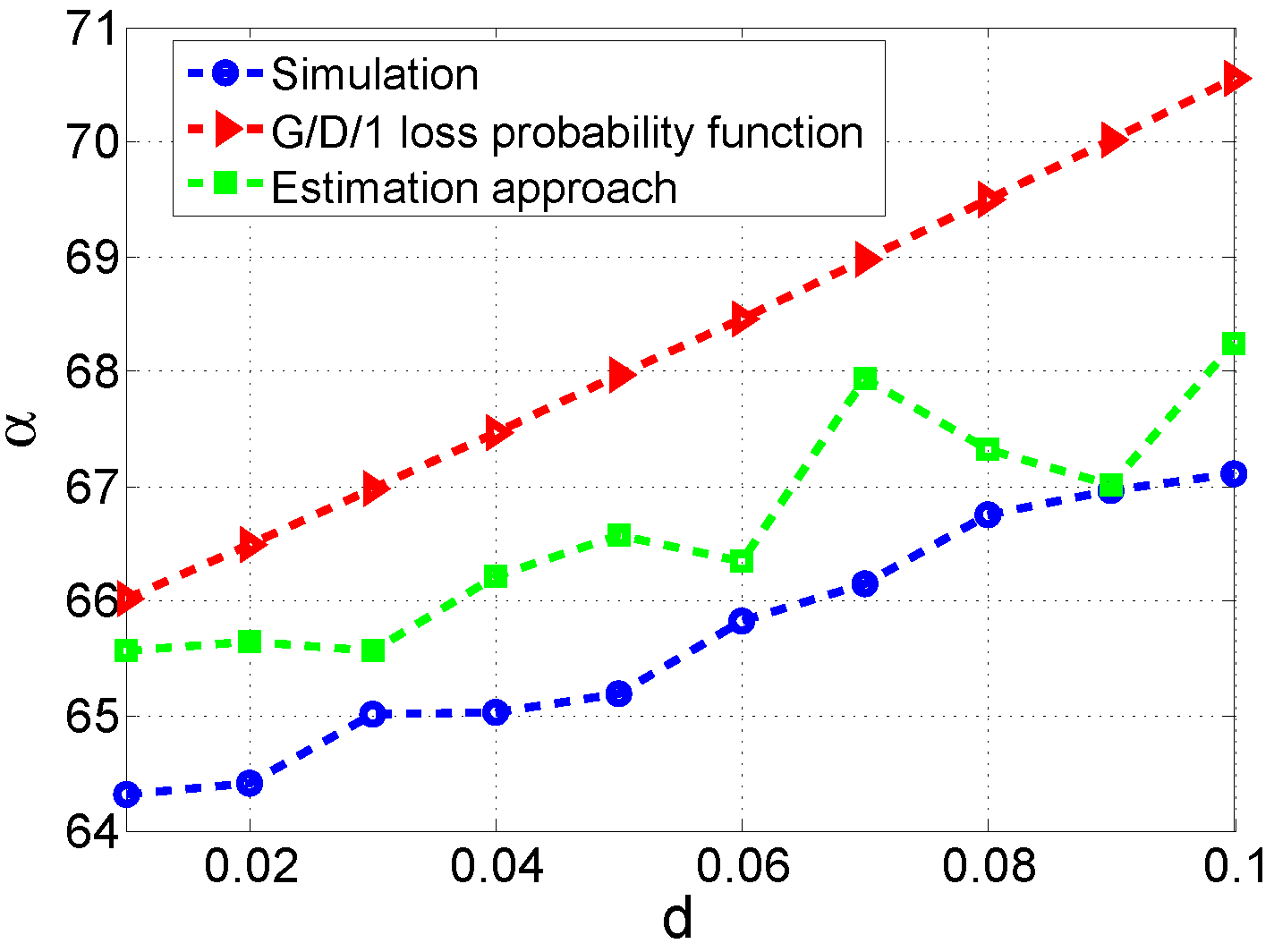}
                \caption{Optimal $\alpha$ versus $D$.}
                \label{fig:62}
        \end{subfigure}
        \begin{subfigure}{0.33\textwidth}
                \includegraphics[width=1.05\linewidth]{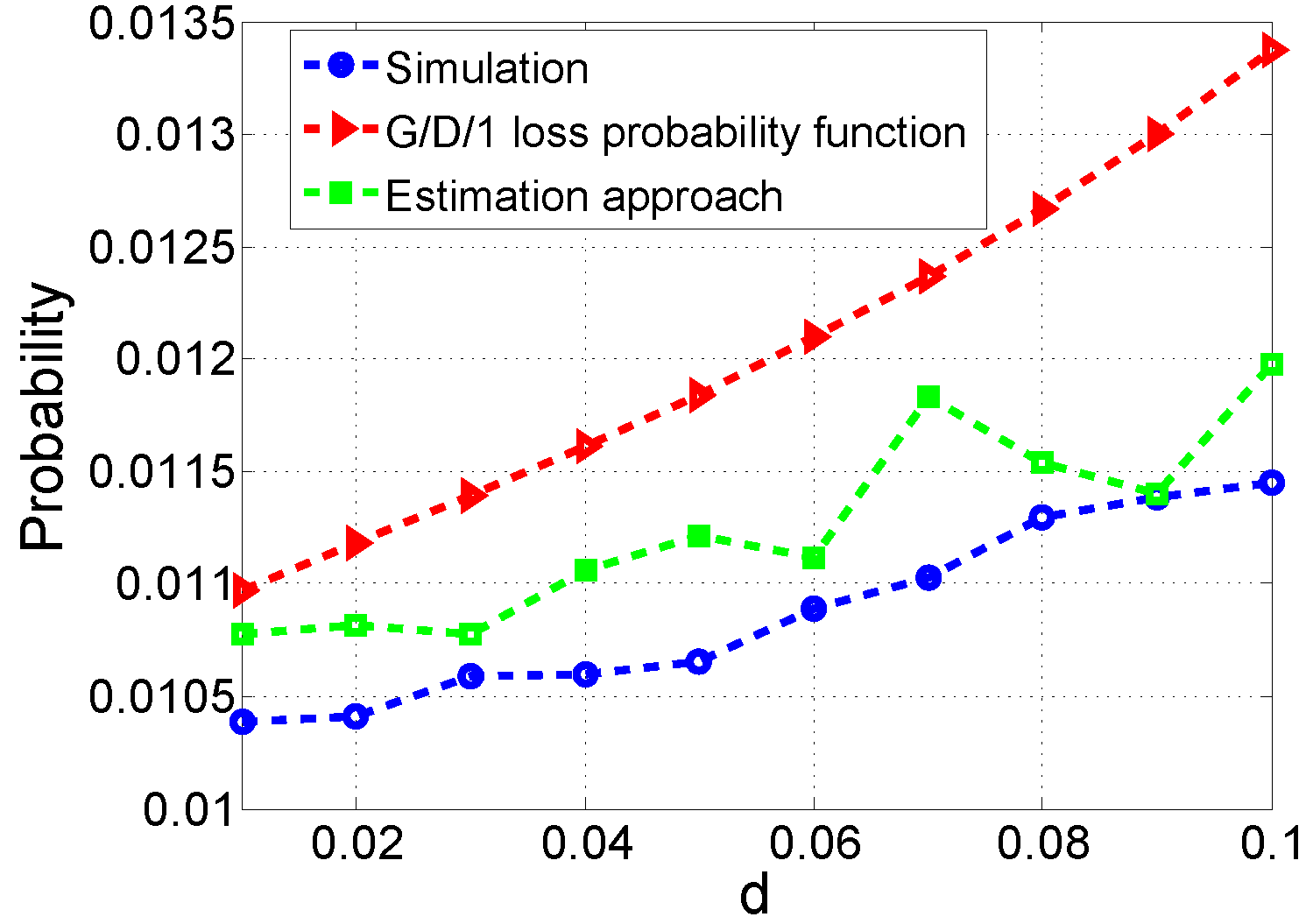}
                \caption{Optimum loss probability versus $D$.}
                \label{fig:63}
        \end{subfigure}%
                \caption{Real data trace based simulations.}
                \label{fig:6}
\end{figure*}
\subsection{Random input}
For the randomly generated input, we assume a total capacity budget of 20 Gigacycles per second. It is also assumed that the average computation workload at shallow cloudlets 1, 2, and 3 is equal to 4, 8, and 6 Gigacycles, respectively. The variance of the input process is also set to one. Moreover, when the input process is modeled by a Gaussian autoregressive (AR) process, the autocovariance is set to $\frac{(0.3)^n}{1-(0.3)^2}$.
For the simulation curves in the figures, the corresponding loss probabilities are calculated by simulations. That is, we neither use the loss probability function formula nor our estimation technique.

Fig.~\ref{fig:2} compares the shape of the loss probability with the that of the proposed upper bound versus $\alpha$ when $D =0.1$ sec.
In particular, Figs.~\ref{fig:2} (a), (b) and (c) show the results for Gaussian AR, Gaussian, and uniform processes, respectively.
Note that the loss probability is defined as the average loss divided by $\sum_{i=1}^{M}\overline{\lambda}_i$. To be comparable with loss probability, the upper bound is also divided by $\sum_{i=1}^{M}\overline{\lambda}_i$ in all the corresponding figures.
As depicted in this figure, the upper bound is minimized almost for the same value of $\alpha$ as the loss probability which confirms the accuracy of the proposed upper bound in terms of optimizing $\alpha$. This figure also evaluates the accuracy of the queue length estimation approach by comparing the upper bound based on this approach with the upper bound based on the simulation.
We do not include the upper bound based on G/D/1 loss probability function~(\ref{equ55}) since this function is valid only for $\alpha\geq\sum_{i=1}^{M}\overline{\lambda}_i$.
Moreover, Fig.~\ref{fig:3} shows the loss probability versus $\alpha$ for Gaussian, Gaussian AR, and uniform input processes when $D=0$.
As shown in Fig.~\ref{fig:3}, in the case of $D=0$, no matter what distribution, the loss probability exhibits a non-decreasing shape versus $\alpha$, which confirms the result of Theorem 1.

Figs.~\ref{fig:4} and~\ref{fig:5} provide the optimization results for different values of $D$ and different input processes. Specifically, Fig.~\ref{fig:4} compares the optimum $\alpha$ of the simulation result with both the G/D/1 loss probability function approach and the queue length estimation approach. Note that the optimum $\alpha$ is increased by increasing $D$ because the queue length at the shallow cloudlets in increased by increasing $D$ and thus, it is more efficient to provide higher capacity at the shallow cloudlets.

Fig.~\ref{fig:5} also compares the same approaches but in terms of the optimum loss probability which is equivalent to the optimum average loss since the loss probability is the average loss divided by constant $\sum_{i=1}^{M}\overline{\lambda}_i$. As depicted in Figs.~\ref{fig:4} and~\ref{fig:5}, while both approaches have high accuracy, the estimation approach yields higher accuracy because the loss probability function is limited to a short range of values of $\alpha$. In other words, the optimum $\alpha$ in the case of G/D/1 loss probability approach is lower bounded by $\sum_{i=1}^{M}\overline{\lambda}_i$. In addition, the better performance of G/D/1 loss probability approach when the input is Gaussian is due to the higher accuracy of function~(\ref{equ55}) for Gaussian input.
Nevertheless, while the estimation approach provides an accurate solution quite close to the simulation, the loss probability of the estimation approach is sometimes lower than that of the simulation. This observation is attributed to the fact that the estimation approach can underestimate the average queue lengths. For example, the queue length estimation method estimates the average queue length as zero ($e_{AQL_i}=0$) for $\alpha\geq\sum_{i=1}^M\overline{\lambda}_i$ while the average queue length based on the simulation is not necessarily zero.

\subsection{Real trace data}
In this section, we simulate the total incoming tasks at shallow cloudlets by the requests made to the 1998 World Cup web site~\cite{Worldcup}
in which we use one hour trend of a sample day for each shallow cloudlet. We also assume that each task requires on average 1 Gigacycles. Fig.~\ref{fig:6} (a) depicts the shape of the loss probability versus $\alpha$ when $D=0$. The loss probability versus $\alpha$ when $D=0$ is a non-decreasing function which confirms the result of Theorem 1 for real trace data as well. Moreover, Figs.~\ref{fig:6} (b) and (c) compare the optimization results, i.e., the optimum $\alpha$ and optimum loss probability, of two proposed approaches with the simulation result. As depicted in these figures, the queue length estimation approach outperforms the G/D/1 approach for the real trace data as well.

\section{Conclusion}\label{sec:conclude}
In this study, we have proposed a new hierarchical capacity provisioning scheme based on accurate queueing analysis. Specifically, we have considered a 2-tier edge computing network architecture consisting of shallow and deep cloudlets, and explored both the case that the network delay between the shallow cloudlets and the deep cloudlet is negligible as well as the case in which the deep cloudlet is located
somewhere deeper in network. Moreover, we have formulated optimization problems for each case and investigated the solution to each problem by using stochastic ordering and optimization algorithms. We have also validated the
performance of our capacity provisioning scheme via extensive simulations.

\end{spacing}

\bibliographystyle{IEEEtran}
\bibliography{ref2}

\end{document}